\documentclass[structabstract]{aa}  
\usepackage{graphicx}
\usepackage{txfonts}
\begin{document}

\title{Investigating variation of latitudinal stellar spot rotation and its 
  relation to the real stellar surface rotation}

\author{H. Korhonen\inst{1,2}
  \and
  D. Elstner\inst{3}
}
\institute{Finnish Centre for Astronomy with ESO (FINCA), University of 
  Turku, V{\"a}is{\"a}l{\"a}ntie 20, FI-21500 Piikki{\"o}, Finland\\
  \email{heidi.h.korhonen@utu.fi}
  \and
  Kiepenheuer-Institut f{\"u}r Sonnenphysik, Sch{\"o}neckstr.\,6,
  D-79104 Freiburg, Germany
  \and
  Astrophysical Institute Potsdam, An der Sternwarte 16, D-14482 
  Potsdam, Germany\\
  \email{delstner@aip.de}
}

\date{Received ; accepted }

 
\abstract
    {}
    {In this work the latitude dependent stellar spot rotation is investigated 
      based on dynamo models, and using similar analysis techniques as in the 
      case of real observations. The resulting surface differential rotation 
      patterns are compared to the known input rotation law used in the 
      calculation of the dynamo model.}
    {Snapshots of the dynamo simulations are used to measure the surface 
      differential rotation. The maps of the magnetic pressure at the surface 
      are treated similarly to the temperature maps obtained using Doppler 
      imaging techniques, and a series of snapshots from the dynamo models are 
      cross-correlated to obtain the shift of the magnetic patterns at each 
      latitude and time point. These shifts are used to study the surface 
      rotation pattern over a wide latitude range at different epochs during 
      the activity cycle, and are compared to the known input rotation law.}
    {Two different rotation laws are investigated, one solar-type law and one 
      with axis distance dependent rotation. Three different dynamo 
      calculations are carried out based on the axis distance dependent law: 
      one with only large scale dynamo field, one with additional strong small 
      scale field and one with weaker small scale field. The surface 
      differential rotation patterns obtained from the snapshots of all the 
      four dynamo calculations show variability over the activity cycle. Clear 
      evolution and variation in the measured surface rotation patterns is 
      seen, but in the models using only the large scale dynamo field the 
      measured rotation patterns are only at times similar to the input 
      rotation law. This is due to the spot motion being mainly determined by 
      the geometric properties of the large scale dynamo field. In the models 
      with additional small scale magnetic field the surface differential 
      rotation measured from the model follows well the input rotation law.}
    {Here, for the first time, the surface differential rotation patterns are 
      investigated in detail based on dynamo calculations. The results imply 
      that the stellar spots caused by the large scale dynamo field are not 
      necessarily tracing the stellar differential rotation, whereas the spots 
      formed from small scale fields trace well the surface flow patterns. It 
      can be questioned whether the large spots observed in active stars could 
      be caused by small scale fields. Therefore, it is not clear that the true
      stellar surface rotation can be recovered using measurements of large 
      starspots, which are currently the only ones that can be observed.}


       \titlerunning{Investigating variation of latitudinal stellar spot 
         rotation}

       \maketitle
%

\section{Introduction}

Differential rotation (DR) is one of the key ingredients in the stellar dynamo 
models. Together with the helical turbulence and meridional flow it is 
responsible for the main features of the solar and stellar magnetic activity 
(see, e.g., R{\"u}diger et al. \cite{rued86}; Choudhuri et al.~\cite{chou95}; 
Brun \& Toomre \cite{brun_toom}). The surface DR of the Sun has been known for 
a long time from observing the rotation of sunspots and other magnetic elements
at different latitudes (e.g., Balthasar et al.~\cite{bal86}). The rotation 
period at the solar equator is approximately 30\% shorter than the period at 
the poles. Helioseismology has revealed that this pattern persists throughout 
the entire convection zone (e.g., Thompson et al. \cite{helio_thom}, Coo et 
al. \cite{helio_schou}), while the radiative core rotates rigidly. Between the 
core and the convection zone there is a transition layer, tachocline, with a 
strong radial shear (Spiegel \& Zahn \cite{tacho}).

Theoretical calculations predict that across the Hertzsprung-Russel diagram DR 
becomes stronger with increasing rotation period and evolutionary status 
(Kitchatinov \& R{\"u}diger \cite{kit_rued99}). In general, models for global 
circulation in outer stellar convection zones predict solar-type DR, where the 
equator is rotating faster than the poles. However, Kitchatinov \& R\"udiger 
(\cite{kit_rued04}) have shown that anti-solar DR, which was suggested by 
observations of several active stars (e.g., Weber \cite{weber07}), could arise 
as a result of intensive meridional circulation.

Stellar differential rotation is usually characterised by the surface shear
$\delta\Omega = \Omega_{\rm eq} - \Omega_{\rm pole}$, where $\Omega_{\rm eq}$
and $\Omega_{\rm pole}$ are the rotation rates at the equator and the poles,
respectively. The surface shear is related to the lapping time,
$t_{\rm lap} = 2\pi/\delta\Omega$. For stars, a surface rotation law of the
form  $\Omega=\Omega_{\rm eq}(1-k\sin^{2}\theta)$ is usually assumed, where
$\theta$ is the latitude. The measurements of surface DR on rapidly 
rotating cool stars is summarised by Barnes et al. (\cite{barnes05}). These 
measurements seem to show an increase in the magnitude of differential rotation
towards earlier spectral types, which is consistent with the theoretical 
calculations of R{\"u}diger~\& K{\"u}ker (\cite{rued_kuek02}). Recently, a 
comparison of the models and observations was carried out by K{\"u}ker~\& 
R{\"u}diger (\cite{kuek_rued08}).

Intriguingly, temporal evolution of the surface DR has been reported for two
young single K stars, AB~Dor and LQ~Hya (see, e.g., Donati et al. 
\cite{donati03}; Jeffers et al. \cite{jeffers07}). Donati et al. 
(\cite{donati03}) hypothesise that these temporal variations could be caused by
the stellar magnetic cycle converting periodically kinetic energy within the 
convective zone into large-scale magnetic fields and vice versa, as originally 
proposed by Applegate (\cite{apple}). 

Measuring stellar differential rotation is not straight forward. Stars are 
faint and apparently small, thus only the closest super giants can be directly 
spatially resolved with the current instruments and methods (e.g., Gilliland~\&
Dupree \cite{gil_dup}; Monnier et al.\,\cite{monnier07}). Currently the best 
way to study stellar surface DR in detail, both the strength and the sign, is 
by using surface maps obtained with Doppler imaging technique (see, e.g., Vogt 
et al.\,\cite{vogt87}; Piskunov et al.\,\cite{pisk90}). One way, similar to 
using sunspots and other magnetic features to study the solar DR, is to use 
several of the surface maps and cross-correlate them to investigate the changes
in the locations of the spots (see, e.g., Barnes et al. \cite{barnes00}; Weber 
et al. \cite{weber05}; K{\H o}v{\'a}ri et al. \cite{kovari07}). The other much 
used way is so-called $\chi^2$-landscape technique in which a DR law is 
implemented into the Doppler imaging code, and many maps are obtained to study 
which parameters give the best solution (e.g., Petit et al.~\cite{petit}; 
Marsden et al.~\cite{marsd06}; Dunstone et al. \cite{dun08}). Stellar surface 
DR has also been studied by combining spot latitude information from Doppler 
images and spot rotation period from contemporaneous photometry (Korhonen et 
al.~\cite{kor07}) and by using Fourier transforms to study shapes of the 
spectral line profiles (see, e.g., Reiners~\& Schmitt~\cite{rei_sch}).

All the methods for measuring stellar surface DR using Doppler imaging results
suffer from the often restricted latitude range the starspots occur on, and 
from possible artifacts in the maps due to systematic errors. These systematics
can for example rise from incorrect modelling of the spectral line profiles. 
Additionally, in $\chi^2$-landscape technique a predefined rotation law is 
assumed. On the other hand, in cross-correlation the time difference between 
the maps is crucial, with too small difference no change has time to occur, and
with too long difference spot evolution due to flux emergence and disappearance
could have occurred. Therefore, it is very demanding to obtain reliable 
measurements of stellar surface differential rotation, and currently it is 
basically impossible to obtain information on the internal rotation of 
stars other than the Sun.

In this study we investigate the surface differential rotation obtained from 
snapshots of dynamo models using cross-correlation methods. These snapshots 
provide a very good temporal base for studying the measured surface 
differential rotation at different times over the stellar activity cycle. The 
obtained surface DR values are compared to the known internal rotation law 
used in the dynamo modelling.

\section{Dynamo model and analysis methods}

\subsection{Model}
\label{model}

The model used here was first developed to theoretically study the flip-flop 
dynamo behaviour. In flip-flop phenomenon the spots concentrate on two 
permanent active longitudes that are separated by 180\degr , and the main spot
activity switches between the longitudes every few years (e.g., Jetsu et al. 
\cite{jetsu93}; Berdyugina \& Tuominen \cite{ber_tuo}; J{\"a}rvinen et al. 
\cite{jarv}). 

The dynamo is modeled with a turbulent fluid in a spherical shell. Two 
different rotation laws are chosen for investigation: one similar to the solar 
one but with a smaller difference between core and surface rotation (L1), and
another one with axis distance dependent rotation (L2), similar to the models 
presented by Moss (\cite{moss04}). The mean electromotive force contains 
an anisotropic alpha-effect and a turbulent diffusivity. 
The nonlinear feed-back of the magnetic field acts on the 
turbulence only. The boundary conditions describe a perfect conducting fluid at
the bottom of the convection zone and at the stellar surface the magnetic 
field matches the vacuum field. The model is discussed in detail by Elstner \& 
Korhonen (\cite{elst_kor}) and Korhonen \& Elstner (\cite{kor_elst}). 

\begin{figure}
  \centering
  \includegraphics[width=4.2cm]{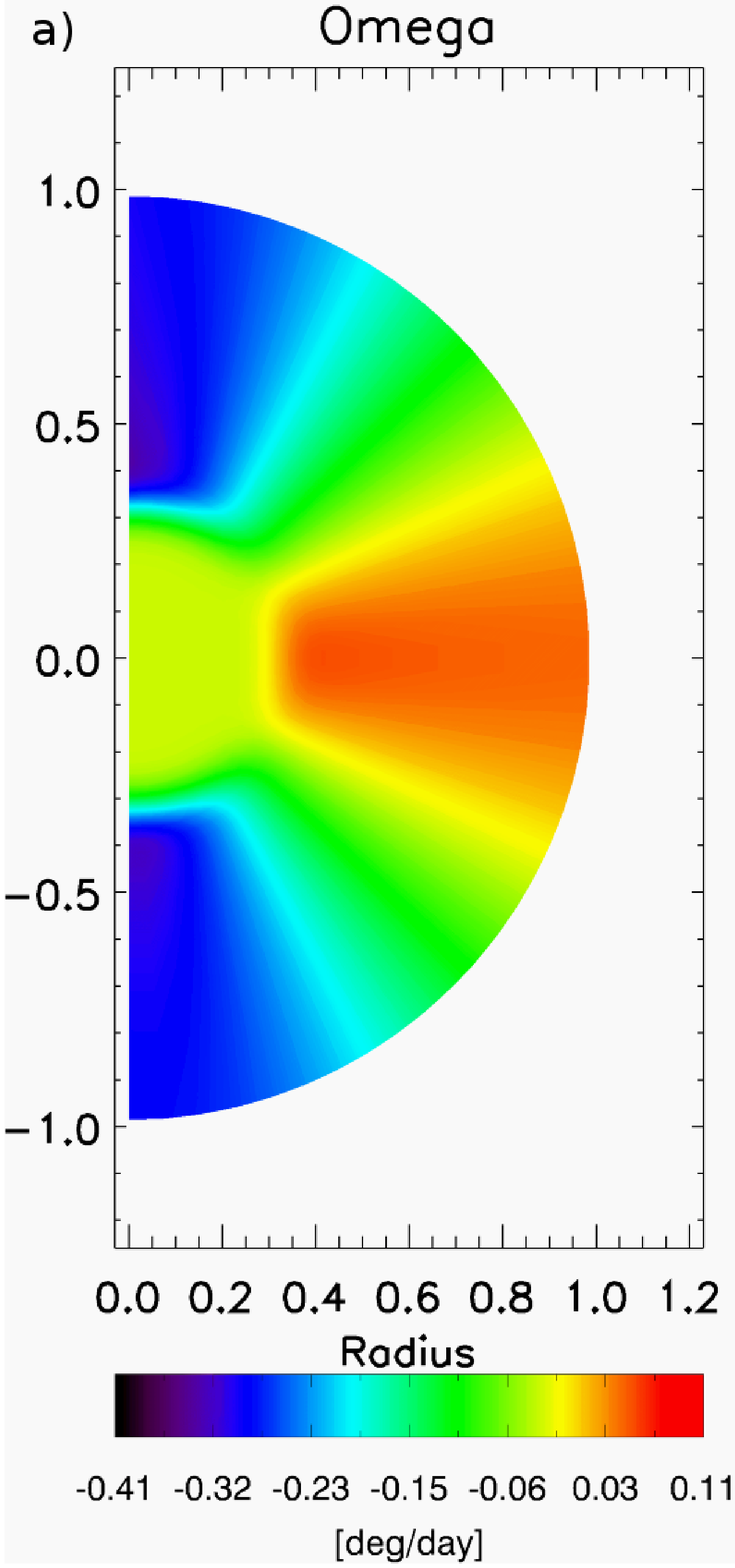}
  \includegraphics[width=4.2cm]{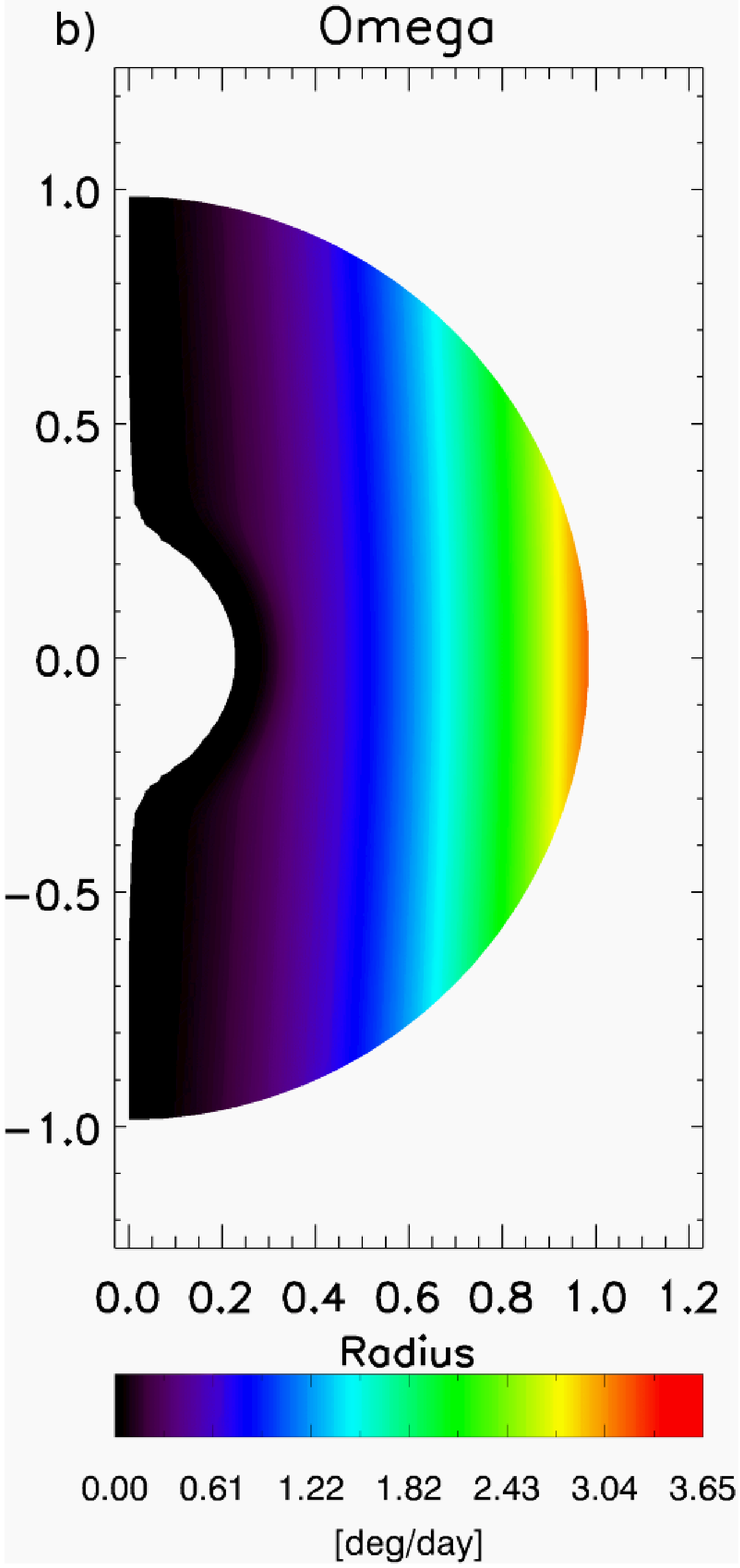}
  \caption{Illustration of the internal rotation laws used in this study. (a) 
    The solar-type rotation law with the equator rotating faster than the poles
    (L1). (b) Axis distance dependent law, where parts furthest away from the 
    axis are rotating the fastest (L2).}
  \label{laws}
\end{figure}

The rotation laws used in this study are illustrated in Fig.~\ref{laws}.
In L1 the internal rotation is solar-type, i.e., the equator is rotating 
faster than the polar regions. The inner boundary of the convective zone is at 
the radius 0.4~R$_{\star}$. L2 is an axis distance dependent rotation law, 
where the parts furthest away from the rotation axis are rotating the fastest. 
The areas close to the axis, i.e., on the surface the polar regions, do not 
rotate at all. In this model the inner boundary of the convective zone is at 
the radius 0.27~R$_{\star}$. The dynamo numbers $C_\alpha=\alpha*R_{\star}/\eta_T$ 
and $C_\Omega=\Delta\Omega*R_{\star}^2/\eta_T$, and the distribution of magnetic 
energy ($e_{0}$ in the axisymmetric mode, $e_{1}$ in the mode with azimuthal 
wavenumber $m=1$, and $e_{n}$ in the higher modes) are given in 
Table~\ref{tab:dynmodel}.

\begin{table}
  \caption{Dynamo models. The name of the model, radius of the inner boundary 
    in stellar radii, dynamo numbers, the distribution of the magnetic energy 
    in different modes, and the cycle period are given.}
\label{tab:dynmodel}
\centering
\begin{tabular}{llllllll}
\hline\hline
name    &       r$_\mathrm{in}$      & $C_\Omega$ &  $C_\alpha$  &  $e_{0}$    &    $e_{1}$     &  $e_{n}$  &             period \\
\hline
              &    [ R$_{\star}$]  &             &         &   [B$_\mathrm{eq}$]      &  [B$_\mathrm{eq}$]& [B$_\mathrm{eq}$] & [years] \\
\hline              
L1    &       0.4    &   184    &       20              &  0.06 &   0.05  &  $6\cdot10^{-5}$ & 6.8  \\
L2    &     0.27    &   250    &       13              &  0.12 &   0.16  &  0.008  &         2.7   \\
L2a  &     0.27   &   250    &       13             &  0.10 &    0.2    &   0.04   &         2.7   \\
L2b  &     0.27   &   250    &       13             &  0.11 &   0.15  &   0.01   &         2.7   \\
\hline
\end{tabular}
\end{table}

\begin{figure}
  \centering
  \includegraphics[width=9cm]{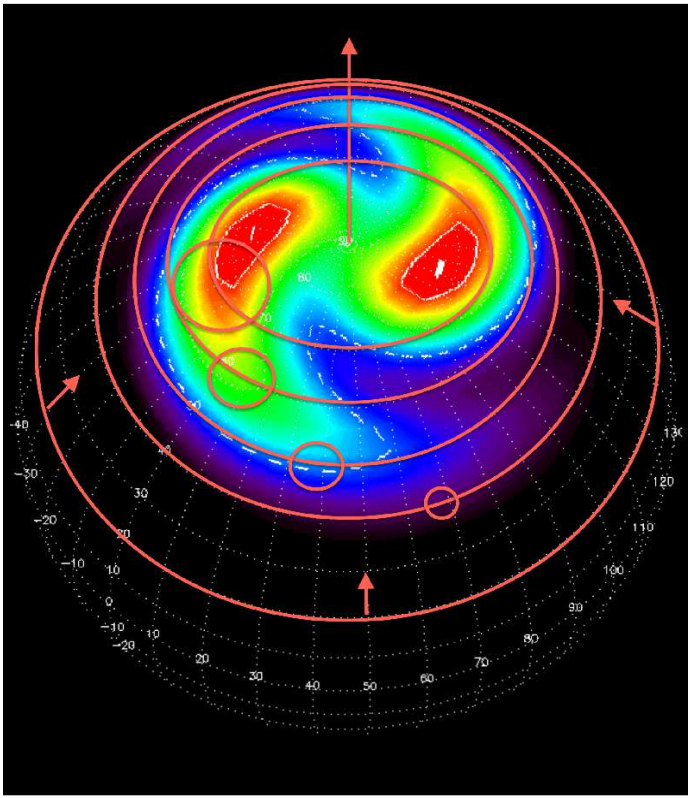}
  \caption{Illustration of the clockwise spot motion (circles) during the 
flip-flop event. Shown is the surface magnetic pressure of the non-axisymmetric
mode. The circles along latitude mark  the maximum of the poleward migrating 
axisymmetric field at different times. Spots appear only at one spiral, where 
the signs of the two modes are equal.  Note, that the stellar rotation and also
the magnetic pattern are anti-clockwise.}
  \label{sketch}
\end{figure}

Both models lead to a mixed field of a non-axisymmetric and an oscillating
axisymmetric mode. The non-axisymmetric modes rotate rigidly with 
$\Omega_m= -0.32\ \mathrm{degrees/day}$ for L1 and $\Omega_m= -2.9\ 
\mathrm{degrees/day}$ for L2. It corresponds to the surface rotation at about 
$55\degr$ latitude for L1 and $65\degr$ for L2. The oscillation period for the 
full cycle of the axisymmetric mode is 6.8 years for L1 and 2.7 years for L2. 
In Fig.~\ref{sketch} the surface magnetic pressure of the non-axisymmetric mode 
for L2 is shown. The schematic rings demonstrate the upward motion of the 
axisymmetric field belt of one polarity during half of a cycle. The small 
circles indicate the spot motion due to the field geometry (in the corotating 
frame of the field pattern). The spots move during one third of a half cycle 
($40\degr$ to $70\degr$) about $100\degr$ opposite  to the stellar rotation. 
Therefore we expect nearly a cancellation of the spot rotation during the 
flip-flop cycle. The detailed behaviour depends on the spiral pattern of the 
field and the migration speed of the axisymmetric field. A similar 
explanation for the differential rotation of active longitudes on the sun was 
already discussed by Berdyugina et al. (\cite{ber_moss}).

In order to study the effect of small scale spots additional simulations were 
run for the model L2, where small poloidal field loops were randomly added into 
the convection zone once per day (laws L2a and L2b). The magnetic energy due to
the additional field injection in model L2a is about half of the energy 
of the axisymmetric field component in the quasi-stationary regime. The fields 
appear as localised spots on the surface, which move with the surface rotation.
With that strong field injection (L2a) the field structure of the global dynamo
mode is nearly hidden by the small scale spots. Therefore in a second run 10 
times weaker field loops (model L2b) were added, which led to a final energy of
the small scale component of only 1/100th of the axisymmetric field. Here the 
global field again dominates the surface spot structure. One could also 
correlate the small scale field injection with the large scale magnetic field, 
but for simplicity a totally uncorrelated field injection was chosen.  

\subsection{Measuring the surface differential rotation}

When using the dynamo models to study the surface differential rotation, the 
temporal resolution can be chosen high enough that standard cross-correlation 
methods can be used. Snapshot maps of the surface magnetic pressure are taken 
from the dynamo calculations and treated as representations of the magnetic 
structures at that time point. These maps can be treated the same way as the 
temperature maps obtained from Doppler imaging, therefore the same techniques 
as in the case of the real observations can be used to analyse the model 
snapshots. Magnetic pressure maps have been taken from 36 time points over the 
activity cycle. These time points are relatively close to each other: separated
by about 50 days in the L1, and by about 18 days with the L2. Examples of 
snapshots for L1 and L2 are shown in on-line 
Figs.~\ref{snap_law1} \&~\ref{snap_law2}, respectively. 

\onlfig{3}{
\begin{figure*}
  \centering
  \includegraphics[width=15.0cm]{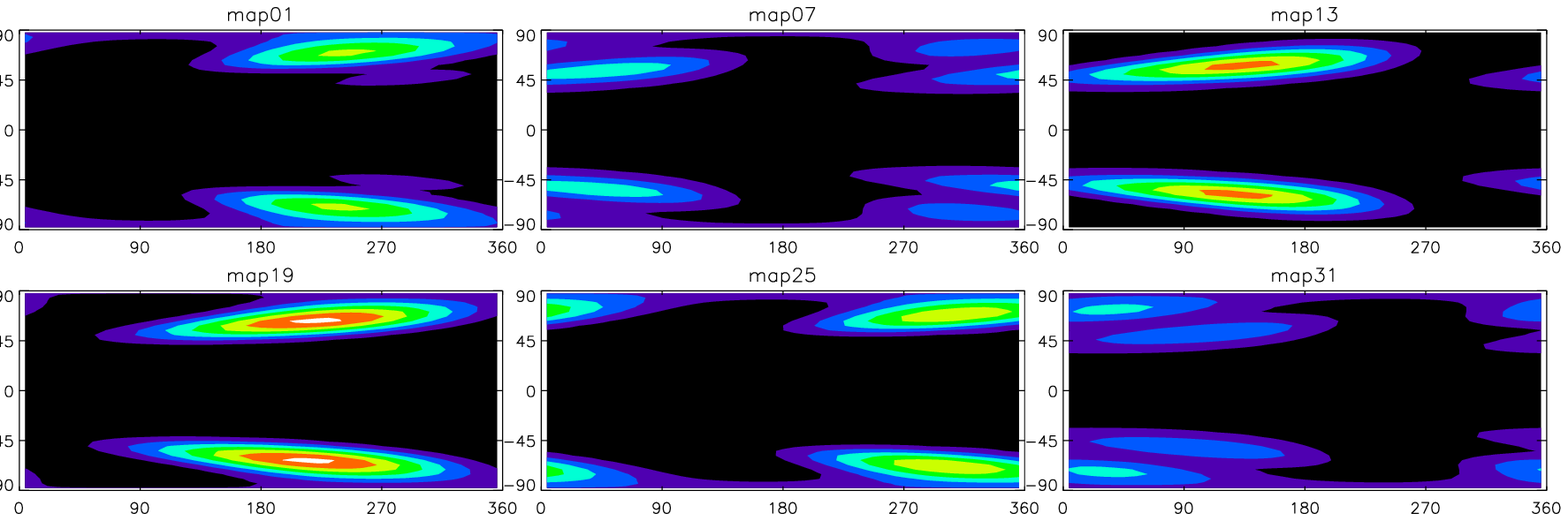}
  \caption{Examples of snapshots from the dynamo model L1. In total 36 
    individual magnetic pressure maps covering a slightly more than one 
    flip-flop event are used in the analysis.}
  \label{snap_law1}
\end{figure*}
}

\onlfig{4}{
\begin{figure*}
  \centering
  \includegraphics[width=15.0cm]{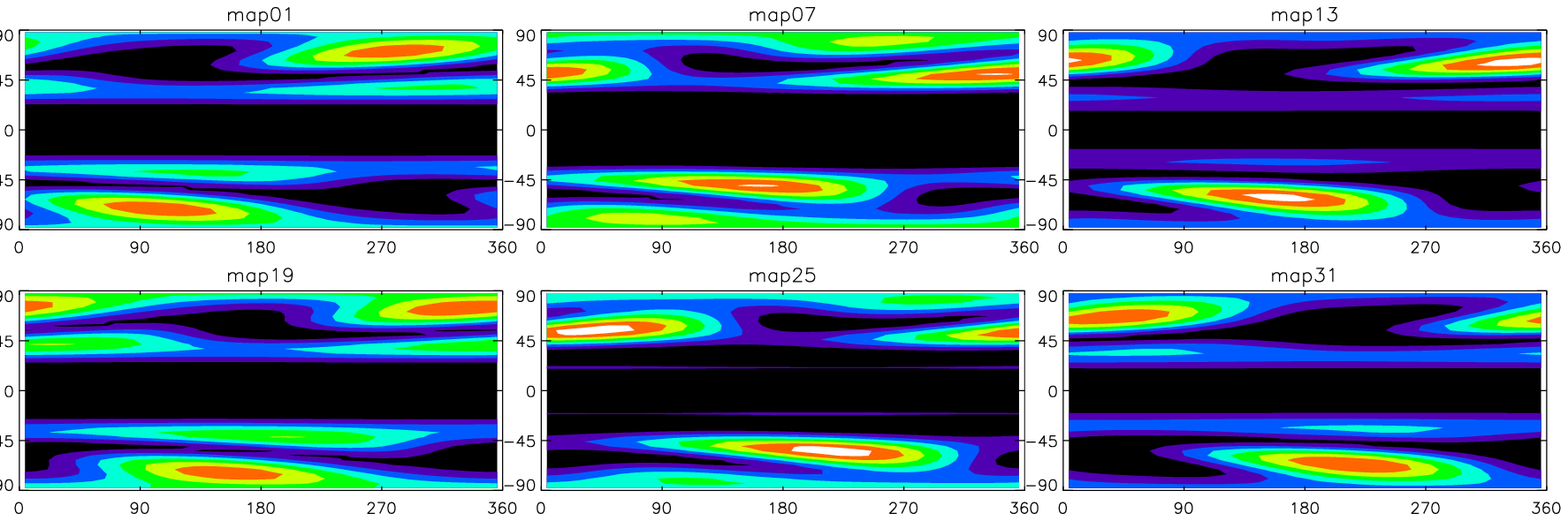}
  \caption{The same as for Fig.~\ref{snap_law1}, except now for model L2.}
  \label{snap_law2}
\end{figure*}
}

In the analysis each time point is cross-correlated with the following one and 
the behaviour is studied from equator to the visible pole. The dynamo model has
a grid of $43\times 43$ surface points. This makes the longitudinal resolution 
of 8.4\degr\ and latitudinal of 4.2\degr. The equator is going through in 
the middle of the 22nd latitude strip. The cross-correlations are obtained for 
each latitude between the latitudes centred at 4.2\degr\ and 83.7\degr. The 
last latitude strip closest to the pole is not used due to the lack of signal. 
The measured shift in degrees/day at each latitude is compared to the expected 
shift obtained from the rotation law used in the dynamo model. The spots are 
migrating in the dynamo simulation, to remove this field migration all the 
measured shifts are normalised to the shift at the lowest latitude used in the 
investigation.

\section{Results}

\begin{figure*}
  \centering
  \includegraphics[width=17cm]{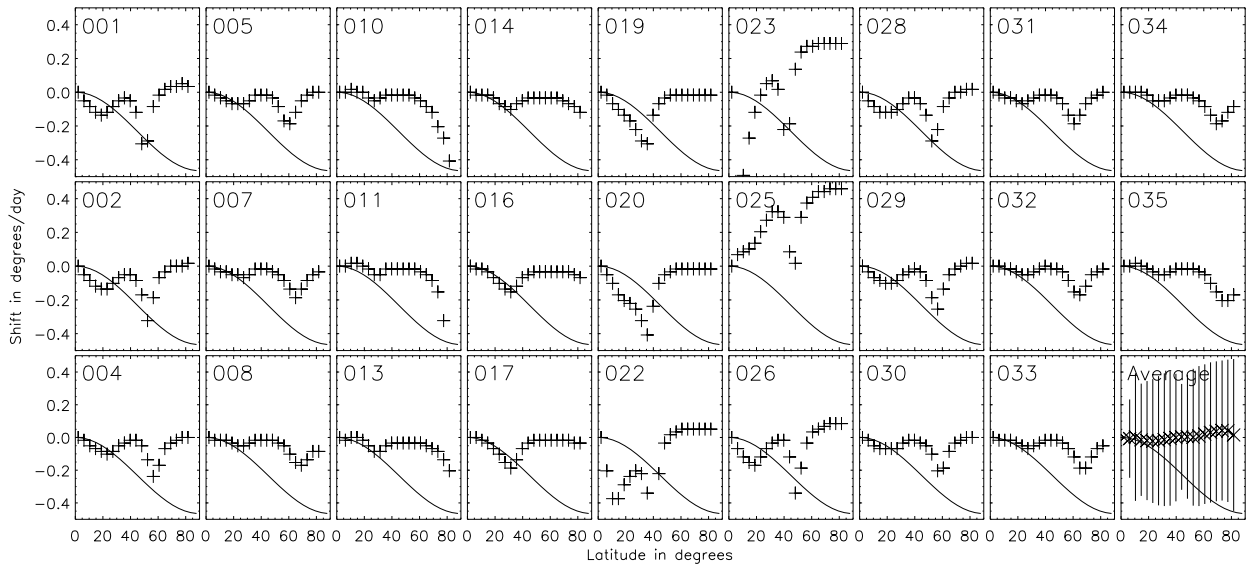}
  \caption {Examples of the cross-correlation results using the 36 
    individual snapshots from the dynamo model L1. The crosses give the shift 
    in degrees/day for each latitude between 4.2\degr\ and 83.7\degr. The 
    dashed line in the plots shows the input rotation law for the dynamo model 
    at the surface. The last plot is the average behaviour obtained from all 
    the snapshots with using the standard deviation of the measurements as the 
    error.}
  \label{CC_solar}
\end{figure*}

Examples of results from cross-correlating the 36 maps for L1 are shown 
in Fig.~\ref{CC_solar}. The plots give the shift in degrees/day for each 
latitude on the visible hemisphere (crosses). The last plot shows the average 
of the measurements from all the cross-correlations, with standard deviation of
the measurements as the error. In the plots the dashed line is the input 
rotation at the stellar surface. 

The behaviour of the measured spot rotation clearly changes from 
cross-correlation to cross-correlation, indicating that the appearing 
surface rotation pattern changes during the activity cycle. Furthermore, the 
measured surface rotation can take numerous different forms: typical solar-type
rotation law (e.g., cross-correlation 11), almost no difference between the 
latitudes (e.g., cross-correlation 01) and even completely anti-solar rotation 
law with the pole rotating faster (e.g., cross-correlation 25). In general it 
is evident that in the case of L1 the measured surface differential rotation 
at the low latitudes often is what one would expect from the models, but at 
higher latitudes the correlation is very poor. 

\begin{figure*}
  \centering
  \includegraphics[width=17cm]{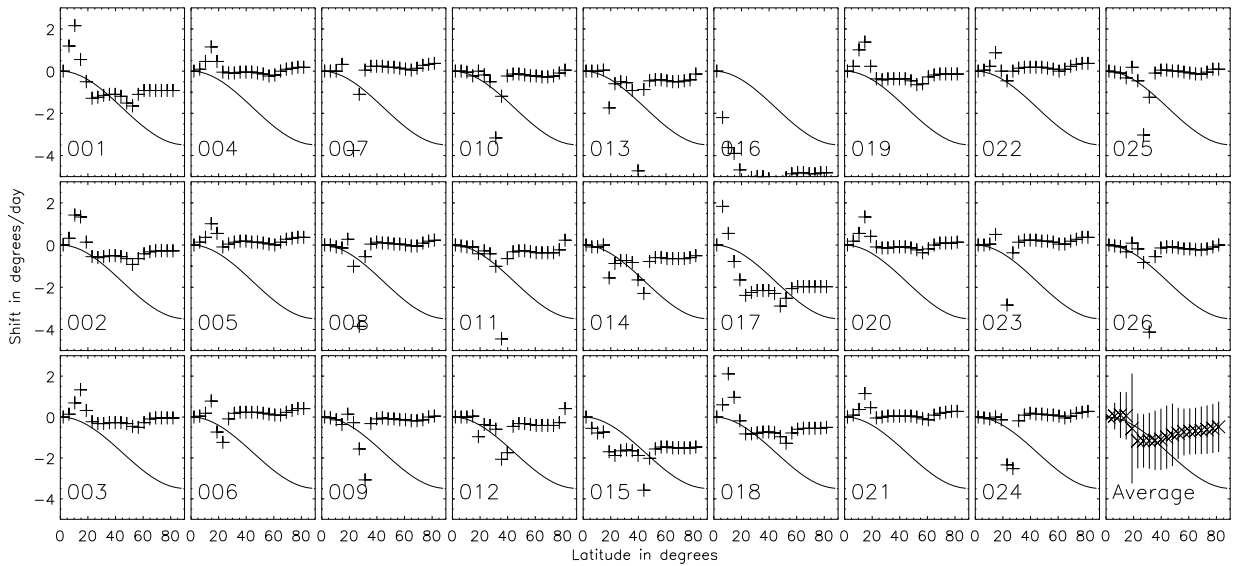}
  \caption{The same as Fig.~\ref{CC_solar}, but using L2. Note that 
    occasionally very large negative shifts are measured and one point can fall
    outside the plotting range.}
  \label{CC_axis}
\end{figure*}

The examples of cross-correlation results from L2, shown in Fig.~\ref{CC_axis},
also exhibit large variations in the surface rotation patterns over the 
activity cycle. In some cases the measured surface rotation at the lower 
latitudes is similar to the input rotation (e.g., cross-correlation~14), at 
times the surface rotation at the equator and at the polar region is similar to
the input rotation, but the exact shape of the rotation curve is different from
the expected (e.g., cross-correlation~16), but in most cases the rotation shows
much less latitudinal variation than expected based on the input rotation.

\section{Discussion}

\subsection{Surface differential rotation patterns over the activity cycle}

\subsubsection{Model L1}

The surface DR measurements for model L1, shown in Fig.~\ref{CC_solar}, 
cover slightly more than one activity cycle, which also includes a flip-flop 
event. Similar phase in the cycle as in the map~1 is reached in the map~28. 
Clear evolution in the surface rotation patterns are seen throughout the cycle.

In general the shape of the surface rotation pattern over the latitude can be 
divided into four parts. Closest to the equator the surface rotation follows 
the internal rotation law, and in most maps three abrupt changes in the surface
rotation occur at higher latitudes. Clear evolution in the latitude at which 
the breaks occur is seen over the activity cycle. When the cycle advances the 
breaks drift towards the higher latitudes. Even at the point that the spots 
move to lower latitude, the breaks continue drifting towards the polar regions.
The highest latitudes for the breaks are obtained just before the time when the
second spot starts to appear in the flip-flop event. When the breaks abruptly 
move towards the equator again, the rotation measurements for the lower 
latitudes become less ordered and appear chaotic. 

In this model the axisymmetric dynamo wave is not only a poleward migration as 
indicated in Fig.~\ref{sketch}. It starts at the bottom of the convection zone 
at $30\degr$ latitude and expands outwards in radius and latitude. For lower 
latitudes the expansion along the non-axisymmetric spiral gives now a counter 
clockwise acceleration. This leads to the qualitative agreement with the 
stellar rotation law near the equator. The effect depends on the expansion 
speed and the pitch angle of the underlying non-axisymmetric field. For the 
rotation law dependent small pitch angle near the equator this is a huge 
effect.  At higher latitudes the spot motion is determined by the field 
geometry and the migration speed of the poleward moving dynamo wave  and 
therefore the rotation is reduced by the clockwise motion of the spot along the
spiral. At the very polar region the wave moves slower and the rotation appears
again faster. This behaviour depends on the phase of the cycle. If the field 
reversal between two cycles lies at mid latitudes ($50\degr$) we see 
the rotation of the non-axisymmetric field, which coincides here with the 
stellar rotation (cf. Cross-correlation 1 to 5 and 26 to 30).

\subsubsection{Model L2}

The measurements of the surface DR using model L2 shown in Fig.~\ref{CC_axis} 
cover almost two activity cycles. The same epoch as in the first 
cross-correlation is again obtained around the time of the 
cross-correlation~18. Evolution in the surface rotation patterns are seen, but 
they are not as clear as in the case of L1.

For most epochs the surface rotation is similar at all the latitudes, showing 
relatively small differential rotation. Often two areas with distinctly 
different rotation behaviour are seen, though. One where the magnetic surface 
patterns lag behind, and one where they rotate faster than the general surface 
rotation. The mainly constant spot velocity in this model goes back onto much 
more clear poleward migration of the oscillating mode because of the different 
shape of the inner rotation law. Here we are missing the equatorward motion of 
the axisymmetric mode and do not see any correspondence to the input rotation 
law. The small positive values sometimes near the equator may be some similar, 
but much weaker, effect as for the model L1. The negative jumps migrating 
poleward are connected to the rigid rotation of the non-axisymmetric dynamo 
mode, seen where the poleward migrating axisymmetric mode is about zero.

The DR measurements for L2 usually show very rigid rotation. During the 
activity cycle the most rigid rotation occurs around the time of the flip-flop
event, and the best fit to the rotation law around the equator and the polar 
region is obtained just before the flip-flop event starts. Still, even at these 
time points the shape of the measured rotation law is very different from the 
shape of the input rotation law. Two deviations from the generally rigid 
rotation are often seen. In many cases a small latitude range shows larger than
normal positive shifts. This area is restricted to latitudes $>$20\degr. Also, 
an area of negative shifts is seen. This area migrates polewards with the 
general migration of the magnetic elements. At the time leading to a flip-flop 
event this area occurs in between the two latitudinal spot belts.

\subsection{Small scale spots versus large scale dynamo spots}

It is interesting to note that the DR measured from the models only including 
large scale dynamo field follows the input DR the best at the latitudes close 
to the equator, and breaks down completely closer to the poles where the 
magnetic pressure is the highest. This could imply that the strongly magnetic 
regions do not follow the differential rotation as well as the less magnetic 
areas.

On the first glance a correspondence of the large scale dynamo field with the 
local differential rotation cannot be expected. Only the non-axisymmetric 
features lead to an observable rotation. The main component here is a large 
scale field with azimuthal wavenumber $m=1$. This field rotates rigidly with a 
value between the minimum and maximum rotation speed. Because of the faster 
rotation at the equator this field is shaped like a leading spiral. Together 
with the oscillating poleward migrating axisymmetric field mode a counter 
clockwise spot motion is introduced, which may be latitude dependent (cf. 
Fig.~ \ref{sketch}). All these effects have no direct connection to the 
differential rotation of the stellar surface. Therefore, the rotation law 
cannot be recovered by measuring the spot motion of the large scale field. On 
the other hand, at least for L1, a good agreement of the cross correlation 
result with the actual rotation law is found at low latitudes. This is a 
consequence of the equatorward extension of the field. Here, a connection to 
the stellar rotation law is obtained through the pitch angle.

Tests were carried out whether the DR patterns can be recovered from models 
including also small scale fields, by adding these fields to the model 
L2. Two different strengths for the small scale fields were used: strong 
field (L2a) and a 10 times weaker field (L2b). For more details on the models 
see Section \ref{model} and Table \ref{tab:dynmodel}.

The examples of snapshots for L2a and L2b are shown in on-line 
Figs.~\ref{snap_law2a} \&~\ref{snap_law2b}, respectively. In the analysis
99 maps are used. In these models, especially in L2a, which has strong small 
scale field, the changes are so rapid that the time step between the snapshots 
is now reduced to approximately 6 days. This means that with the 99 snapshots 
we cover more or less the same time period as with the 36 snapshots earlier.

\onlfig{7}{
\begin{figure*}
  \centering
  \includegraphics[width=15.0cm]{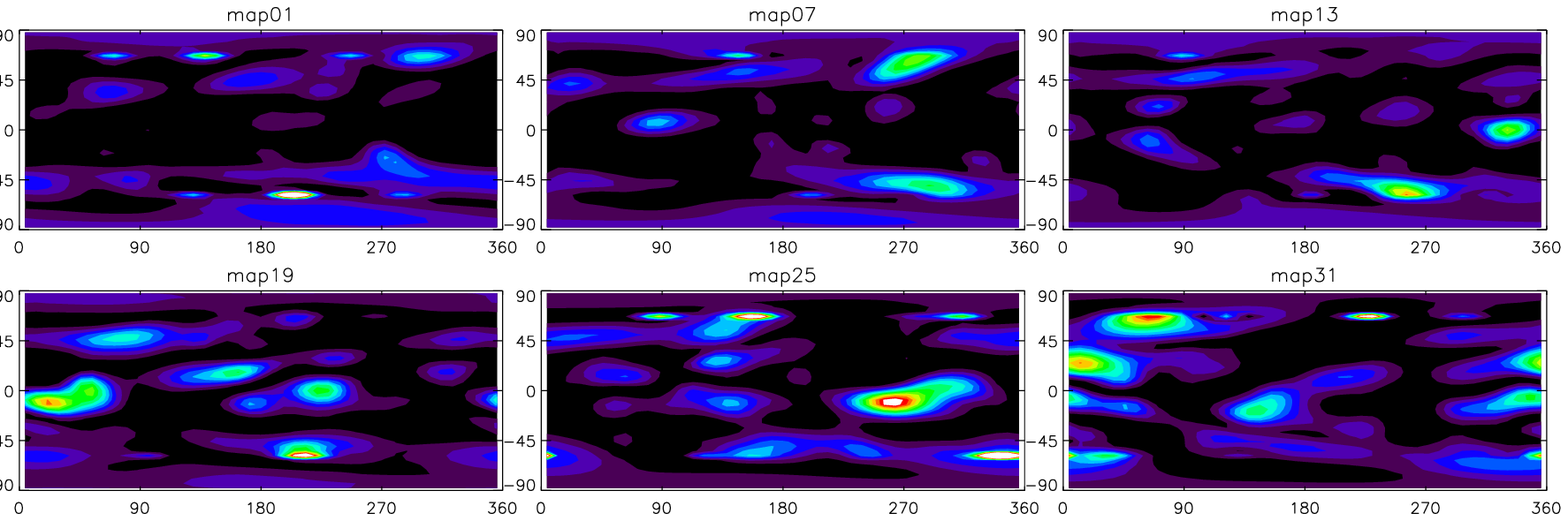}
  \caption{Examples of the snapshots from the dynamo model L2a. In total 99 
    magnetic pressure maps are used in the analysis.}
  \label{snap_law2a}
\end{figure*}
}

\onlfig{8}{
\begin{figure*}
  \centering
  \includegraphics[width=15.0cm]{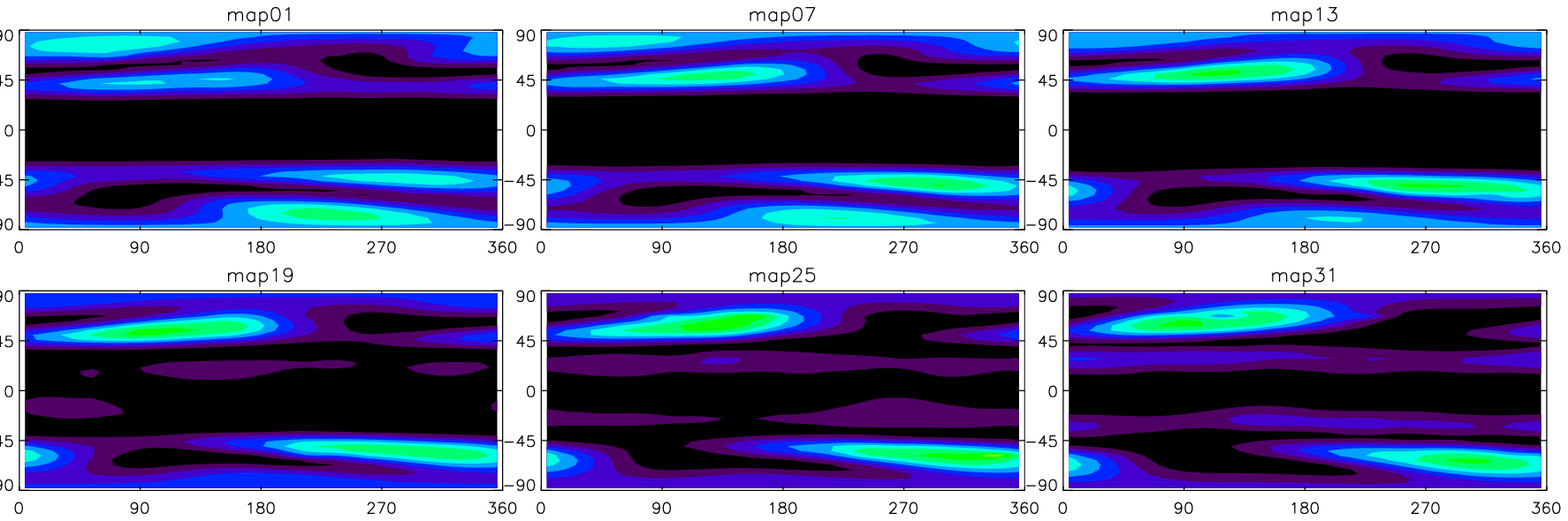}
  \caption{The same as for Fig.~\ref{snap_law2a}, except now for model 
    L2b.}
  \label{snap_law2b}
\end{figure*}
}

\begin{figure*}
  \centering
  \includegraphics[width=17cm]{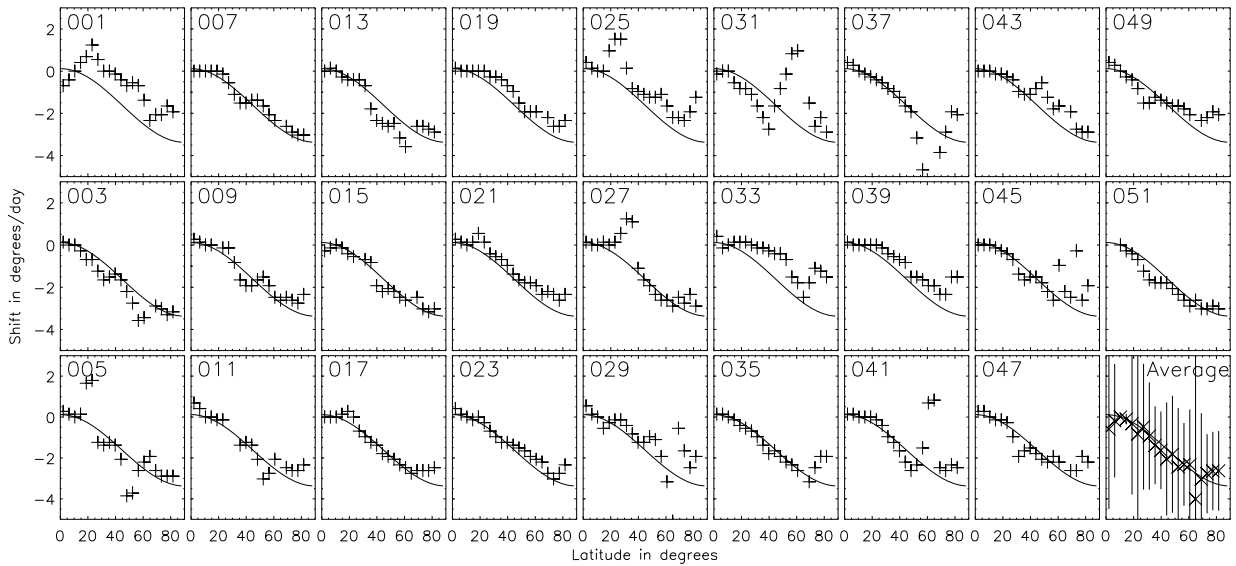}
  \caption {Examples of the cross-correlation results using all the 99 
    individual snapshots from the dynamo model L2a. The symbols are the same as
    in Fig.~\ref{CC_solar}. The last plot is the average behaviour obtained 
    from all the 99 snapshots with using the standard deviation of the
    measurements as the error. Latitude 12.6$^{\circ}$ is used as the reference 
    latitude.}
  \label{CC_law2a}
\end{figure*}

\begin{figure*}
  \centering
  \includegraphics[width=17cm]{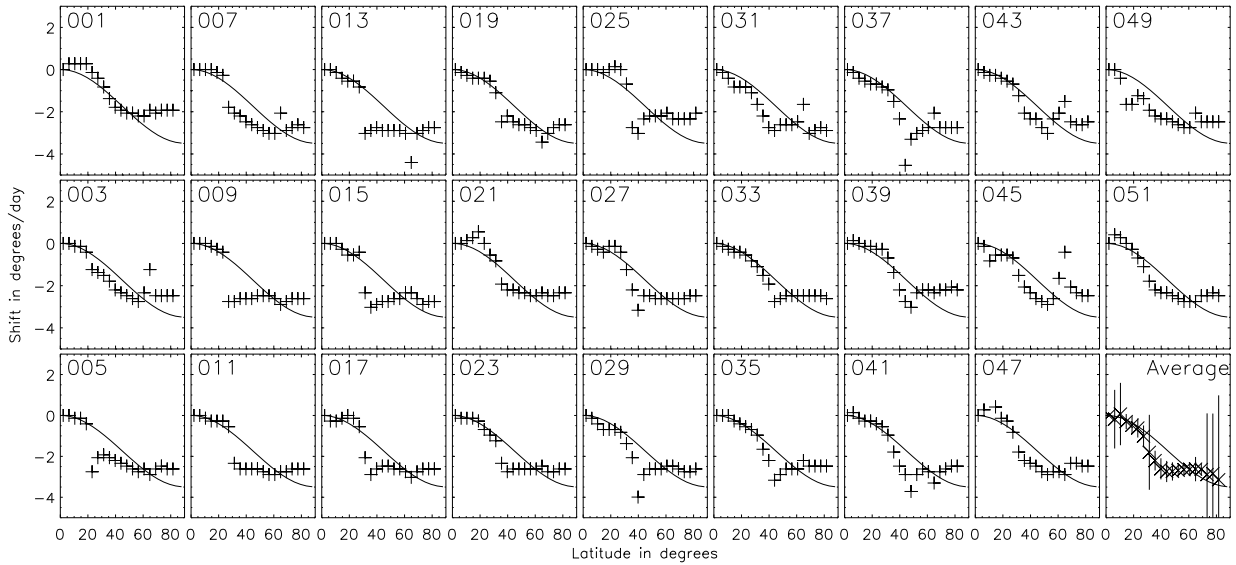}
  \caption {As for the Fig. \ref{CC_law2a}, but now for L2b and using 
    latitude 4.2$^{\circ}$ as the reference latitude.}
  \label{CC_law2b}
\end{figure*}

Examples of the cross-correlation results from  L2a are shown in 
Fig.~\ref{CC_law2a}. The cross-correlation is done for 99 snapshots of magnetic
pressure, but only every other cross-correlation curve for the time period 
covering a bit more than one flip-flop cycle is shown in the plot. Here, the 
latitude 12.6$^{\circ}$ is used as the reference latitude due to the anomalous 
shifts occasionally measured at the lower latitudes. As can be seen, now the 
recovered surface DR follows well the input rotation. Still, some time points 
of anomalous rotation are seen, but, as shown by the average rotation, in 
general the rotation is well reproduced. 

The results for L2b are shown in Fig.~\ref{CC_law2b}. Now the latitude 
4.2$^{\circ}$ is again used as the reference latitude. With a weaker small scale
field of the model  L2b the results are between the results from the 
original L2 and from L2a, in which stronger small scale field was used.
Again, the general rotation law is recovered, but the high latitudes start to 
again show almost rigid rotation.

Addition of the small scale fields to the dynamo model enables us to recover 
the input rotation law. But it remains an open question if the observable
large spots can be attributed to  small scale magnetic features. Recently, 
Lehtinen et al. ~(\cite{lehtinen11}) found active longitudes rotating 
significantly slower than the star, which according to the current 
investigation, favors spots as signs for the large scale field.  

\subsection{Comparison to observations}

The results presented here clearly show that the rotation law obtained from 
dynamo models with large scale fields breaks at the latitudes where most of the
spots occur. 

A literature search was carried out to investigate whether or not 
this phenomenon is also seen when cross-correlating Doppler images. One 
significant difference between the model calculations used here and the real 
Doppler images, is that when cross-correlating the models there is always 
signal, even if the signal is weak, i.e., not resulting in a starspot, whereas 
in Doppler images the latitudes devoid of spots only result in noise in the
cross-correlation. 

The literature search reveals that at times the cross-correlation result 
reproduces very well the $\sin^{2}$-type surface rotation law within 
observational errors, which at times can be very large. A relatively good fit 
is seen for example in RS~CVn-type binaries IM~Peg (Weber~et 
al.~\cite{weber05}) and IL~Hya (Weber~\& Strassmeier \cite{web_str98}), young 
single K dwarfs AB~Dor (Donati~\& Collier Cameron~\cite{don_cc}) and LQ~Hya 
(K{\H o}v{\'a}ri~et al.~\cite{kovari04}), and for a post T~Tauri star 
RX~J1508.6\ $\pm$ 4423 (Donati~et al.~\cite{donati00}). On occasions the 
$\sin^{2}$-law has the same general shape as what is obtained from 
cross-correlating latitude strips in the Doppler images. For example the shifts
measured for a late-type single star PZ~Tel follow well the general shape of 
the $\sin^{2}$-law, but there are latitudes, especially around the equator, 
where the measurements deviate significantly from the $\sin^{2}$-law (Barnes et
al. \cite{barnes00}). Also for V1192~Ori the general shape of the rotation 
law is reproduced, but deviations are seen at the higher latitudes (Strassmeier
et al. \cite{str_etal03}). On KU~Peg the latitudinal rotation rates form a 
complex double peaked pattern, where the equator and the latitudes around  
45$^{\circ}$ rotate faster than the general rotation (Weber \& Strassmeier 
\cite{web_str01}; Weber et al.~\cite{weber05}). There are also examples of 
targets where Doppler images obtained from different spectral lines give 
slightly different latitudinal rotation behaviour, e.g., $\sigma$~Gem 
(K{\H o}v{\'a}ri~et al.~\cite{kovari07}) and UZ~Lib (Vida~et al.~\cite{vida07})

One has to also note that the results on cross-correlating Doppler images do 
not commonly extend to higher latitudes than 60$^{\circ}$. Often this is 
due to the polar spot, which is seen in many of the maps. This constant feature
would not result in shifts in spot configuration between two consecutive 
Doppler images, and thus it would not give any signal in the cross-correlation.
Still, this feature is often large, and the strongest spot on the surface,
i.e., analogous to the maximum magnetic pressure in the dynamo models, and 
exactly the location where the rotation behaviour is seen to deviate from the 
input rotation. Thus, it is very difficult to conclude based on the Doppler 
images whether or not the observations also show the behaviour where the 
surface rotation in strongly magnetic features deviates from the general 
rotation.

\section{Conclusions}

Starspots are generally used as tracers of the stellar surface differential 
rotation. If starspots are direct signs of the global stellar magnetic field 
this assumption does not seem to be valid, because in this case the spot motion
would mainly be determined by the geometric properties of the large scale 
dynamo field. Based on the dynamo calculations presented here, the surface 
differential rotation can only be recovered by following the spot motion of 
small scale spots frozen into the flow.

In the dynamo models presented here the usual assumption of a positive 
alpha in the northern hemisphere is made, implying a poleward migration 
of the axisymmetric dynamo wave. This gives together with the leading spiral of
the non-axisymmetric field geometry a strongly reduced or even counter rotating 
spot motion. This could possibly explain the reported detections of 
anti-solar differential rotation in some stars (e.g., Vogt et 
al.~\cite{vogt99}; Weber et al.~\cite{weber05}). Nevertheless, for equatorward 
drifting dynamo waves, as in the Sun, the spot motion will more or less 
coincide with the surface differential rotation. In our model L1 it coincides 
even quantitatively in the equatorial region. This may be due to a connection 
of the strength of the differential rotation to the pitch angle of the 
non-axisymmetric mode. The fact that mainly solar type rotation laws have been 
observed (see, e.g., Barnes et al.~\cite{barnes05}) would imply that the 
dynamos in general show an equatorward  drift. Further models with more solar 
type dynamos are needed. In case that the observed spots are more local 
phenomena we have a direct correspondence of the spot motion to the stellar 
rotation law. This was nicely reproduced using the cross-correlation method.

Here, stellar surface differential rotation was for the first time investigated
in detail based on dynamo calculations. The following conclusions can be drawn 
from this study:

\begin{enumerate}
\item Cross-correlation is a powerful method to follow differential rotation 
  using starspots as tracers, if the time difference between the maps being 
  cross-correlated is appropriate for detecting the differential rotation 
  induced changes. 
\item Spots due to the large scale dynamo field are not necessarily tracing the
  real surface differential rotation.
\item Equatorward drift of the dynamo field gives at least qualitatively an 
  agreement with the differential rotation, as was seen for the model 
  with a solar type rotation law (L1). 
\item The differential rotation obtained from the spot motion varies during the
  dynamo cycle.
\item Spots caused by small scale magnetic fields are good tracers of the 
  stellar differential rotation.   
\item Comparing direct tracers of differential rotation (e.g., analysing shapes
  of spectral lines) and obtaining the spot motion from cross-correlation
  can probably clarify the nature of the spots, i.e, whether they are caused by
  small scale field or are signs of the large scale dynamo field.
\end{enumerate}

\begin{acknowledgements}
      With this article we would like to remember Prof. Ilkka Tuominen who died 
      in March 2011. He will be missed by his colleagues, but also remembered 
      by his work on stellar activity and dynamos. Part of this work was 
      supported by the German \emph{Deut\-sche For\-schungs\-ge\-mein\-schaft, 
        DFG\/} project number KO~2320.
\end{acknowledgements}

\end{document}